\documentclass[aip,pof,graphicx]{revtex4-1}

\usepackage{amsmath}
\usepackage{graphicx}
\usepackage{color,cancel}

\usepackage{natbib}

\begin{document}

\title{The effect of the Basset history force on particle clustering in Homogeneous and Isotropic Turbulence} 

\author{S.~Olivieri}
\affiliation{DICCA, University of Genova, Via Montallegro 1, 16145 Genova, Italy}

\author{F.~Picano}
\affiliation{Department of Industrial Engineering, University of Padova, Via Venezia 1, 35131, Padova, Italy}
\affiliation{SeRC and Linn\'e FLOW Centre, KTH Mechanics, SE-100 44 Stockholm, Sweden}

\author{G.~Sardina}
\affiliation{SeRC and Linn\'e FLOW Centre, KTH Mechanics, SE-100 44 Stockholm, Sweden}
\affiliation{Department of Meteorology, SeRC (Swedish e-Science Research Centre), Stockholm University, 
106 91 Stockholm, Sweden}
\affiliation{Facolt\'a di Ingegneria e Architettura, UKE Universit\'a Kore di Enna, Enna, 94100, Italy}

\author{D.~Iudicone}
\affiliation{Stazione Zoologica A. Dohrn,
villa Comunale, Naples, Italy}

\author{L.~Brandt}
\affiliation{SeRC and Linn\'e FLOW Centre, KTH Mechanics, SE-100 44 Stockholm, Sweden}

\date{\today}

\begin{abstract}
We study the effect of the Basset history force on the dynamics of
small particles transported in homogeneous and isotropic turbulence and show
that this term, often neglected in previous numerical studies, 
reduces the small-scale clustering typical of inertial 
particles. The contribution
of this force to the total particle acceleration is, on average, responsible for about 10\% of the total acceleration and particularly relevant during 
rare strong events. 
At moderate density ratios, i.e. sand or metal powder in water,  its presence alters the balance of forces determining the particle acceleration.

\end{abstract}

\pacs{}

\maketitle 

{\bf Introduction.}
Turbulent flows with dispersed phases can be found in several engineering applications and natural phenomena.
On one side, we may recall sprays of droplets~\cite{domingo2005dns} and solid particles~\cite{mcdonald2005direct}, typical of many combustion systems, whereas
rain formation~\cite{falkovich2002acceleration} and plankton\cite{reynolds:book} are examples of natural flows. 
Preferential concentration in specific flow events 
affects the sedimentation rate 
of heavy particles~\cite{wang:1993} and the plankton dynamics in the ocean~\cite{reynolds:book}. 
The solid phase transport properties
depend on the particle inertia, or better the particle relaxation time $\tau_p$, and on the particle to fluid density ratio $R$~\cite{toschi-rev:2009}. 
Particles with high density ratio and relaxation time $\tau_p$ of the order of the flow dissipative time scale (Kolmogorov time) $\tau_K$ 
exhibit intense small-scale clustering~\cite{toschi-rev:2009}.
The particle velocity field may  also exhibit caustics, i.e. different particle velocities at the same position~\cite{falkovich2002acceleration,bec2010intermittency}. 
Both these phenomena crucially influence the collision rate and make difficult to provide an accurate estimate of it. 

The relevance of the forces determining the particle acceleration varies with the particle density ratio and size. The Lagrangian equation governing the motion of small particles dispersed in a flow has been formulated by Maxey and Riley~\cite{maxey:1983}.
The particle acceleration is determined by the Stokes Drag, due to the viscous forces acting on the particle, the Pressure Gradient, 
related to the pressure difference on the opposite sides of the particle,
the Added Mass
due to the  {relative acceleration between the particle and the fluid}, Gravity and the 
History force due to the unsteady diffusion of vorticity in the boundary layer around the 
particle.  %
{Elgobashi and Truesdall \cite{elgo92}
and Armenio and Fiorotto \cite{armenio:2001} numerically investigated  the relevance of these different terms on particles dispersed in decaying homogeneous isotropic turbulence and turbulent channel flow. 
Both studies report that  for all cases explored the Basset History term is 
not negligible with respect to the other forces. 
Armenio and Fiorotto \cite{armenio:2001} also found that the Pressure Gradient governs the
particle dynamics for density ratios of order one, while the Stokes Drag is dominant at high density ratio ($R \sim100$). Nonetheless, significant differences on the particle dispersion in
turbulent channel flows when including the Basset force are not shown.} Note however, 
that more complex phenomena, i.e.\ turbophoresis~\cite{reeks1983transport}, arise in wall-bounded flows due to the strong spatial
inhomogeneity~\cite{sardina2012wall}.
The large part of numerical studies on inertial particles do not account for the Basset History term, see Ref.~\onlinecite{toschi-rev:2009} for a recent review, 
an approximation usually considered quite safe for particles with high density ratio $R$.  
In stratified turbulence the vertical dispersion of particles with $R\sim1$ is found to be attenuated by about $15\%-20\%$ when the Basset history term is considered in the dynamics~\cite{aartrijk:2010a}. 
Recently, Daitche and T{\'e}l~\cite{daitche2011memory} 
have studied the effects of the Basset history term on particle transport in two-dimensional unsteady
wakes and
shown that the particle clustering and the caustic formation are strongly reduced, {as} reported also for particles 
in two-dimensional chaotic flows~\cite{PhysRevE.88.042909}.  

 {The modeling of the history force is crucial in our study: here we neglect finite Reynolds number effects on the  convolution Kernel. During the '90s a series of studies examined inertial effects on the history force. In particular, Mei and Adrian\cite{meiadr} found that at long times and finite Reynolds numbers, the kernel decays faster, $\sim t^{-2}$, than the Basset kernel, $\sim t^{-1/2}$,  a scaling obtained assuming a sinusoidal variation of the uniform flow around the particle and a constant particle Reynolds number, based on the mean uniform velocity. This makes the model hardly applicable to Lagrangian particle tracking in turbulence. 
Lovalenti and Brady\cite{lovbra} obtained a more general expression for the history force at finite Reynolds number. The kernel decays exponentially and the results of Mei and Adrian are recovered for particles accelerating from rest. The model is  computationally too expensive  and we therefore choose to consider the classic Basset kernel, making sure it is valid for the parameters of our simulations. Ling et al.\cite{lingetal} provide a general overview on this debate with useful criteria to select the proper Kernel (Fig.\ 3 of that paper). We adopt their criterion based on the flow Kolmogorov time scale $\tau_K$ and the viscous-unsteady time scale $\tau_{vu}$, function of the particle/flow  parameters. If $\tau_K \ll \tau_{vu}$ then diffusive Basset kernel should be used. Our data satisfy this condition for all cases except for the highest  $R = 1000$, and $\mathit{St}_\mathrm{K}=\frac{\tau_\mathrm{p}}{\tau_\mathrm{K}}=1$, where $\tau_K/\tau_{vu}$ is about 1 
 and  the clustering attenuation reported below is likely to be overestimated. 
However, we chose to present these data in the following as  an upper limit for the effects of the Basset kernel history force on 
particle dynamics.}

The main aim of the present work is thus to quantitatively characterize the effect of the Basset history term on particle clustering and its contribution to the instantaneous particle acceleration in three-dimensional 
homogenous and isotropic turbulence. 
We will show that the small-scale clustering of inertial particles  is always reduced by the Basset history term. 
In addition, we observe that at $R=\mathcal{O}(1-10)$ the importance of the forces determining the instantaneous particle acceleration changes
when the Basset history is considered. 

{\bf Methodology.}
We  
use an Eulerian-Lagrangian approach to describe the dynamics of the fluid and particle phases.
The carrier flow is governed by the incompressible Navier-Stokes equations
where we denote $\mathbf{u}(\mathbf{x},t)$ the fluid velocity field, $\rho_\mathrm{f}$ the fluid density, 
and $\nu$ is the fluid kinematic viscosity.
We consider rigid spheres smaller than the relevant hydrodynamic scales, i.e.\ the Kolmogorov length $\eta=(\nu^3/\epsilon)^{1/4}$ 
(with $\epsilon$ the average energy dissipation), and assume {small} particle Reynolds number,
$\mathit{Re}_\mathrm{p}={r_{p} |V_0-U_0|}/{\nu} \ll 1$
with   $V_0$ and $U_0$ the particle and fluid velocities,  and $r_p$ the particle radius.  {Dealing with small particles,  finite-size
effects, e.g.\ Fax{\'e}n correction or Saffman lift, can be safely neglected with particles behaving as material 
points.
In addition, dilute conditions are considered in order to neglect particle-particle interactions and the back-reaction
 on the carrier phase (one-way coupling), see Ref.~\onlinecite{balachandar-rev:2010} and reference therein 
 for related discussions.}
 {Considering the previous assumptions and neglecting gravitational effects}, 
the particle velocity obeys the Maxey-Riley equation~\cite{maxey:1983}:
\begin{multline}
 {
\underbrace{\frac{\mathrm{d}\mathbf{V}}{\mathrm{d}t}}_{a_\mathrm{p}} =\underbrace{ \frac{\mathbf{u}-\mathbf{V}}{\tau_\mathrm{p}}}_{a_\mathrm{SD}} + \underbrace{\frac{\rho_\mathrm{f}}{\rho_\mathrm{p}}\frac{\mathrm{D}\mathbf{u}}{\mathrm{D}t}}_{{a}_\mathrm{PG}} + 
+ \underbrace{\frac{1}{2}\frac{\rho_\mathrm{f}}{\rho_\mathrm{p}}\,(\frac{\mathrm{D}\mathbf{u}}{\mathrm{D}t}-\frac{\mathrm{d}\mathbf{V}}{\mathrm{d}t})}_{{a}_\mathrm{AM}}
+\underbrace{\sqrt{\frac{9}{2\pi}\frac{\rho_\mathrm{f}}{\rho_\mathrm{p}}\frac{1}{\tau_\mathrm{p}}} \int_{-\infty}^{t}\frac{1}{\sqrt{t-\tau}}\frac{\mathrm{d}}{\mathrm{d}\tau}(\mathbf{u}-\mathbf{V})\,\mathrm{d}\tau}_{{a}_\mathrm{Ba}}}
\label{MR_acc}
\end{multline}
where 
$\mathbf{u}(\mathbf{X}(t),t)$ is the flow velocity sampled at the particle position,
$\rho_\mathrm{p}$ is the particle density,
$\tau_\mathrm{p}=({2}/{9}) ({r_\mathrm{p}^{2}}/{\nu}) ({\rho_\mathrm{p}}/{\rho_\mathrm{f}})$
is the particle response time. 
The acceleration terms on the RHS of eq.~\eqref{MR_acc} are the Stokes Drag  {($a_\mathrm{SD}$)}, Pressure Gradient  {($a_\mathrm{PG}$)}, 
Added Mass  {($a_\mathrm{AM}$)} and Basset History  {($a_\mathrm{Ba}$)},  {while ${a}_\mathrm{p}$ is the particle acceleration.}
The particle position $\mathbf{X}(t)$ is given by ${\mathrm{d}\mathbf{X}}/{\mathrm{d}t}=\mathbf{V}(t)$.

Fixing the flow conditions, i.e.\ the Reynolds number,  {and neglecting gravity,}
the particle dynamics depends only on  {two} dimensionless parameters: the 
density ratio $R=\rho_\mathrm{p}/\rho_\mathrm{f}$, {and} the Stokes number
$\mathit{St}_\mathrm{K}=\frac{\tau_\mathrm{p}}{\tau_\mathrm{K}}$,
with  $\tau_\mathrm{K}=\nu/\eta$ the Kolmogorov time.
 
The Navier-Stokes equations are solved by means of Direct Numerical Simulations using a classic pseudospectral method on a triperiodic cubic domain, whereas particles are analyzed in a Lagrangian framework.
A quadratic interpolation provides the values of flow quantities at the particle position, those needed to solve the Maxey-Riley equation.
The particle velocity and trajectory are evolved in time by using a third-order Adams-Bashfort scheme.

The computation of the Basset history term, the integral term in the eq.~\eqref{MR_acc}, 
presents the greatest numerical challenge as it requires the full history of the particle acceleration.
To evaluate this force at a reasonable cost, we use the method recently proposed in Ref.~\onlinecite{hinsberg:basset}, briefly introduced here for completeness.
The integration is split in two parts,  denoted as window and tail.
The first consists of a numerical integration over the interval $[t-t_\mathrm{win},t]$, considering $N_\mathrm{w}$ previous steps.
The singularity of the kernel function $1/\sqrt{t-\tau}$ for $t=\tau$ 
is dealt with by the 
trapezoidal rule 
\begin{equation}
\begin{split}
\mathbf F_\mathrm{Ba,win}(t) &\approx \frac{4}{3} C_\mathrm{Ba} \sqrt{\Delta t} \, \mathbf{b}_0 
             + \sum_{n=1}^{{N_\mathrm{w}}-1} \frac{4}{3} C_\mathrm{Ba} \sqrt{\Delta t} \biggr[ (n-1) \sqrt{n-1} - 2 n \sqrt{n} + (n+1) \sqrt{n+1} \biggr] \mathbf{b}_n + \\
             &+ C_\mathrm{Ba} \sqrt{\Delta t} \biggr[ \frac{4}{3} ({N_\mathrm{w}}-1) \sqrt{{N_\mathrm{w}}-1} + (2-\frac{4}{3}{N_\mathrm{w}}) \sqrt{{N_\mathrm{w}}} \biggr] \mathbf{b}_{N_\mathrm{w}} 
\end{split}
\end{equation}
where $C_\mathrm{Ba} = 6r_\mathrm{p}^{2}\rho_\mathrm{f}\sqrt{\pi\nu}$,
$\Delta t=t_\mathrm{win}/{N_\mathrm{w}}$ the time step
and $\mathbf{b}_n$ the discretized value of $\mathbf{b}(\tau) = \frac{\mathrm{d}}{\mathrm{d}\tau}(\mathbf u - \mathbf V)$.
The remaining history $[-\infty,t-t_\mathrm{win}]$ is approximated using recursive exponential functions leading to lower computational costs still keeping high accuracy.
As suggested in Ref.~\onlinecite{hinsberg:basset}, we chose $N_\mathrm{w}=5$ for all the simulations 
as this value gives a good compromise between accuracy and computational effort.

All simulations were performed using $288^3$ grid points on a cubic domain of side $\mathcal{L}=2\pi$.
The turbulence is characterized only by the Taylor Reynolds number $\mathit{Re}_\lambda=u'\lambda/\nu \simeq 136$, 
with
$\lambda=\sqrt{\epsilon/(15\nu u'^2)}$ and  $u'$ the root-mean-square of the fluid velocity fluctuations.  {The ratios between the characteristic turbulent scales are $\mathcal{L}/\lambda=15$ and $\mathcal{L}/\eta=370$.}

\begin{figure}
\centering
\includegraphics[width=.42\textwidth]{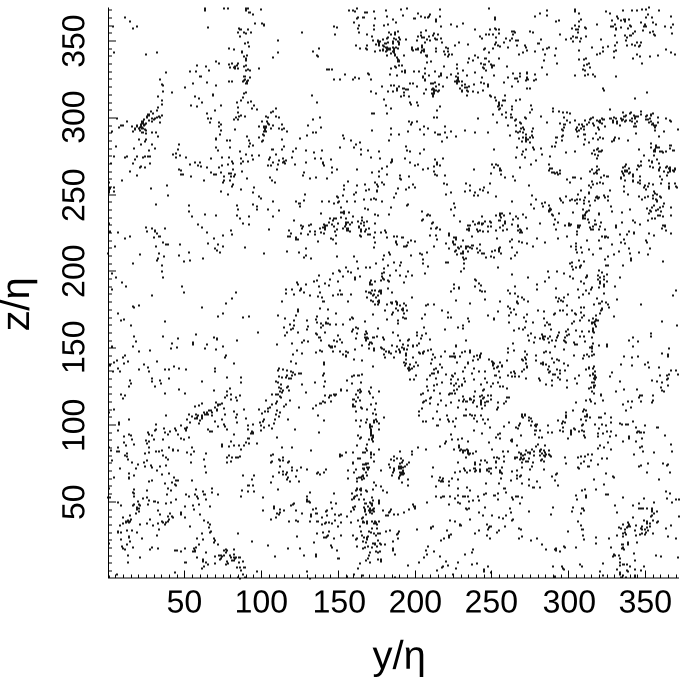}~~~~
\includegraphics[width=.42\textwidth]{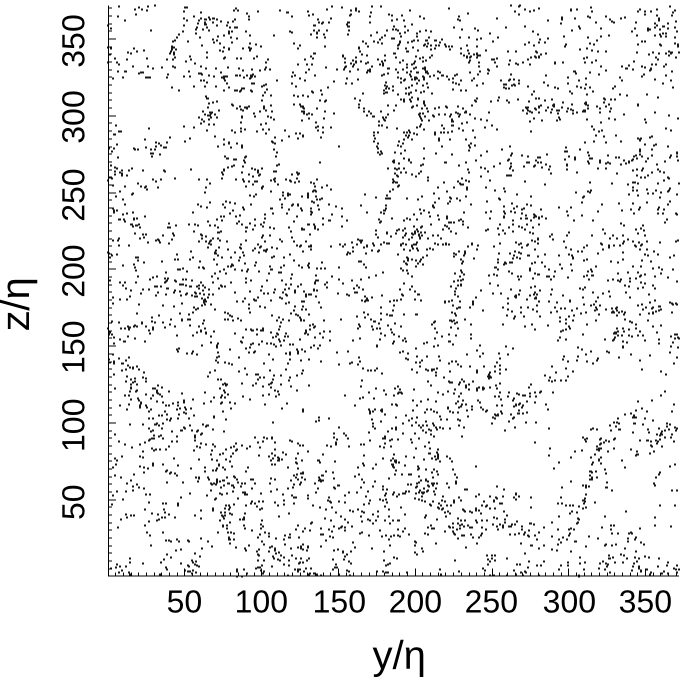}
{\scriptsize \put(-315,170){\bf (a) without Basset term}}
{\scriptsize \put(-120,170){\bf (b) with Basset term}}
\caption{Screenshots of particles distribution at $R=10$, $St_{{K}}=1$.  
(a) Results from a DNS without the Basset history term. (b) Results including 
all terms of eq.~\eqref{MR_acc}, thus also  the Basset
force. \label{fig:pos}}
\end{figure}

We consider different combinations of density ratio $R$ and Stokes number $St_K$: in particular at
 $R=1$: $St_K=0.01, \, 0.1$, at $R=10$: $St_K=0.01, \, 0.1, \, 1$, at $R=1000$: $St_K=0.01, \,  0.1, \,  1$. 
  $R=1$ and $R=10$ well reproduce the behavior of small solid particles in liquids, while $R=1000$ is typical of aerosol/droplets in gases. The
 Stokes numbers are selected to avoid particles larger than the hydrodynamic lengths, so within the limits of our model  {(the ratio $r_p/\eta$ ranges from 0.0067 to 0.67)}. For each parameter set, we performed simulations with and without the Basset history force for comparison.
We have simulated the unladen fluid phase until reaching the  fully developed turbulent regime, when the particles 
are introduced with a random spatial distribution and the velocity of the fluid at the same position.
We evolve the particle-fluid system for a time $T \simeq 700 \, \tau_\mathrm{K}$ 
 and save snapshots
 every $0.7 \, \tau_\mathrm{K}$ time units
 in order to compute statistics.

{\bf Results.}
Snapshots of the particle position from the simulation with $R=10$ and $St_{{K}}=1$ are displayed in figure~\ref{fig:pos}. 
Results in (a) are obtained without the Basset history terms as often assumed in previous numerical studies on
particle-laden turbulent flows. Small-scale clustering characterizes the particle distribution \cite{toschi-rev:2009}: clusters 
and void regions large enough
to be clearly appreciated at first sight. When the Basset history ( {Ba}) term  is included, the particle segregation appears to be less intense, i.e.\  the Basset history term acts to smear out the clusters, as also recently observed 
in chaotic bi-dimensional flows~\cite{daitche2011memory,PhysRevE.88.042909}.

\begin{figure}
\centering
\includegraphics[width=.48\textwidth]{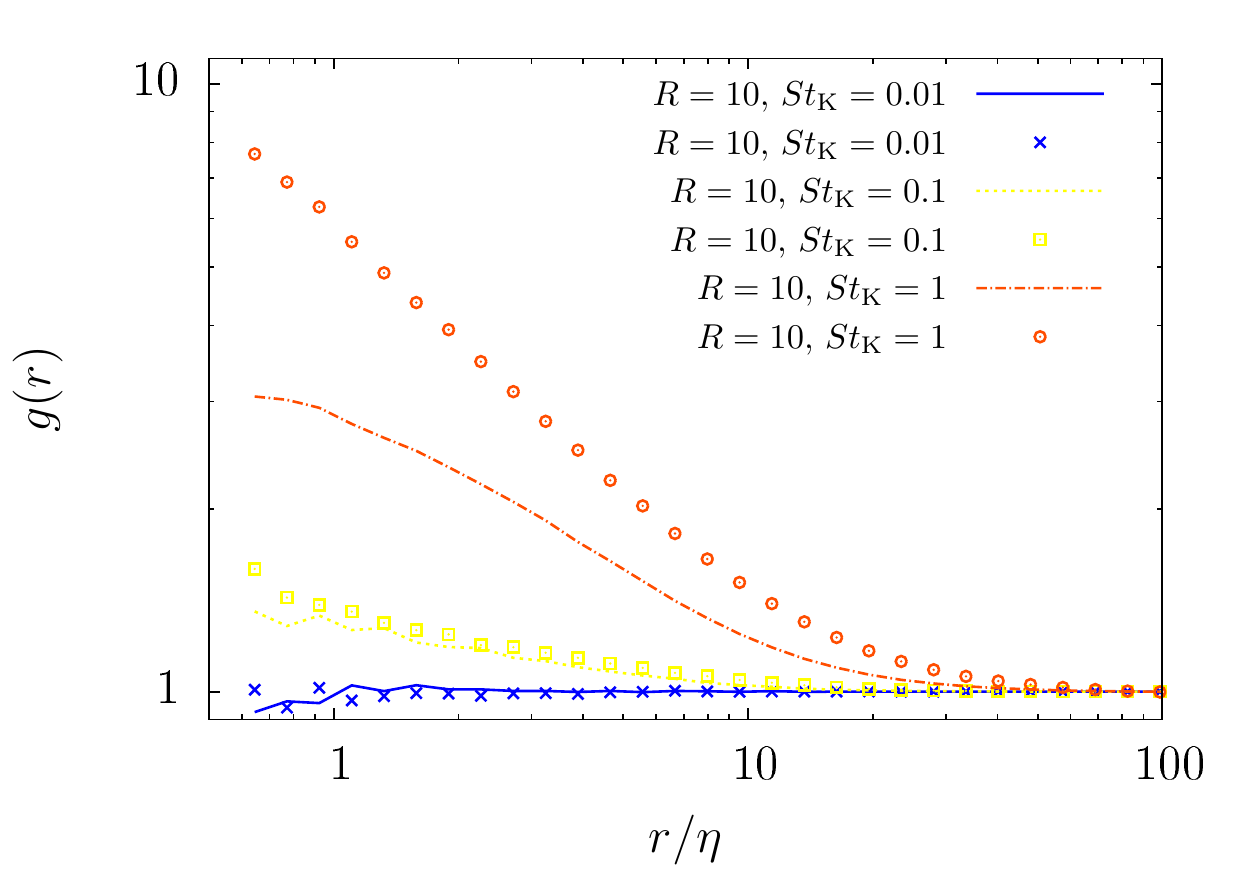}~~~~
\includegraphics[width=.48\textwidth]{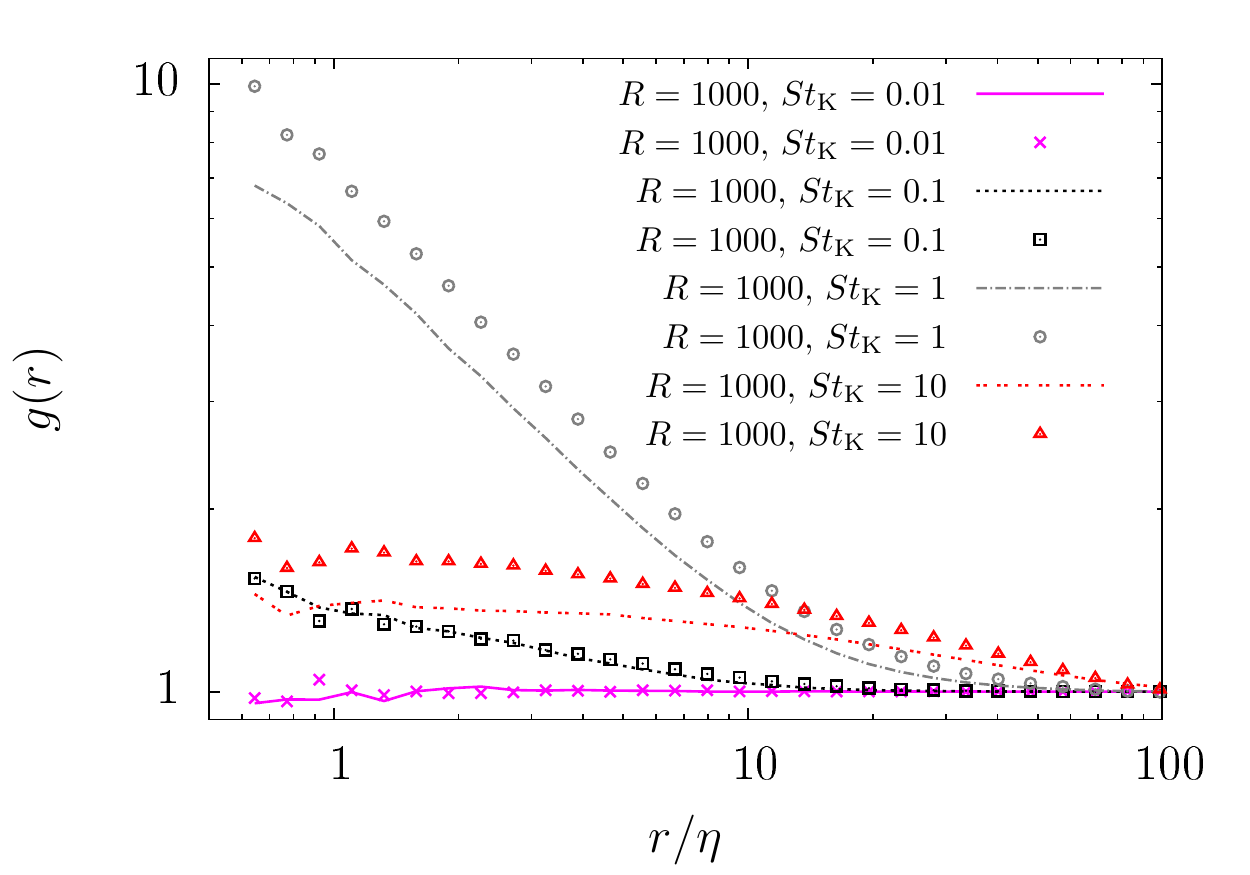}
{\put(-395,110){(a)}}
{\put(-185,110){(b)}}
\caption{The radial distribution function $g(r)$ versus particle distance $r/\eta$  for (a)  $R=10$, and (b) $R=1000$.
Results including the Basset term are displayed by solid lines, results without Basset with symbols.
\label{fig:par_rad}
}
\end{figure}

The clustering intensity can be quantified by the radial distribution function (RDF), $g(r)$, measuring  the probability to find a particle pair at distance $r$ normalized with 
that of a purely random Poissonian arrangement. The RDF for some of the most representative cases ($R=10$ and $R=1000$)
is reported in Fig.~\ref{fig:par_rad} where we compare data obtained with (lines) and without
(symbols) considering the  {Ba}.
For both density ratios, we find the highest levels of small-scale clustering when $St_K=1$, while the accumulation is weaker for the other Stokes numbers considered, in agreement with previous findings \cite{toschi-rev:2009}. 
For all cases,
the effect of the  {Ba} term is to weaken the clustering, confirming the visual impression given in figure~\ref{fig:pos}.   
Particles with $R=1$ or tiny $St_K$ do not show clustering, 
consistently with previous investigations~\cite{PhysRevE.86.035301}, and this does not change including the  {Ba} term.

The relative importance of the different forces appearing on the RHS of equation~\eqref{MR_acc} is here studied by varying $R$ and $\mathit{St}_\mathrm{K}$.
We aim to determine the relevance of the Basset history term to the total particle acceleration by comparing simulations with and without this term.  
Since the mean values of the whole acceleration is null in homogeneous and isotropic flows we will examine the p.d.f. (\emph{probability density function}) of the different 
acceleration sources. 
In particular, we consider  the p.d.f. of each term on the RHS of the Maxey-Riley equation~\eqref{MR_acc} 
divided by the total particle acceleration,
\begin{equation}
 1 = \frac{a_\mathrm{SD}}{a_\mathrm{p}} + \frac{{a}_\mathrm{PG}}{{a}_\mathrm{p}} 
 + \frac{{a}_\mathrm{AM}}{{a}_\mathrm{p}} + \frac{{a}_\mathrm{Ba}}{{a}_\mathrm{p}}
\end{equation}
 {where the different} terms denote vector components and not the modulus.
By computing the p.d.f. of each of the quantities appearing above, a clear indication of the dominant contributions will be determined.
Values around 1 will indicate a component that alone determines the overall instantaneous particle acceleration.

\begin{figure}
\centering
\includegraphics[width=.48\textwidth]{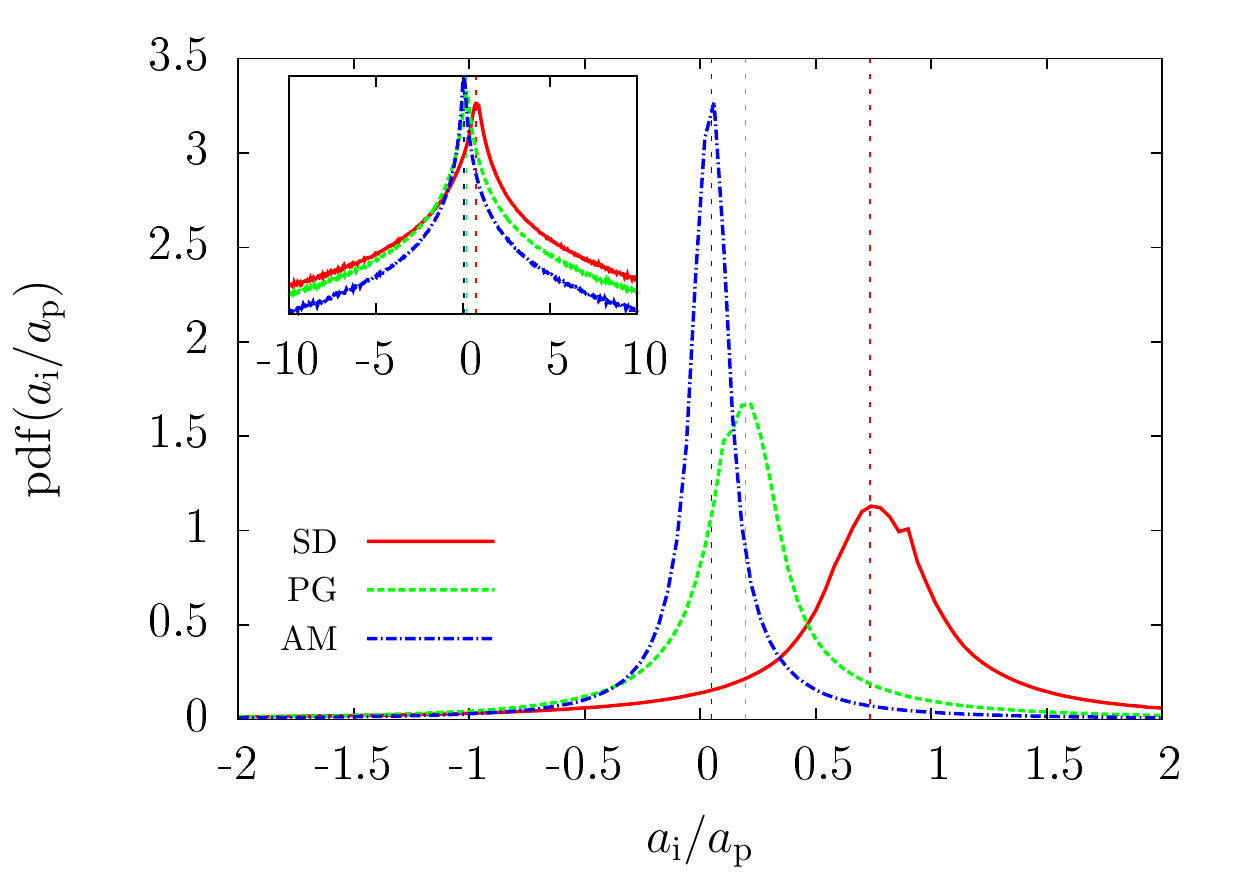}~~~~
\includegraphics[width=.48\textwidth]{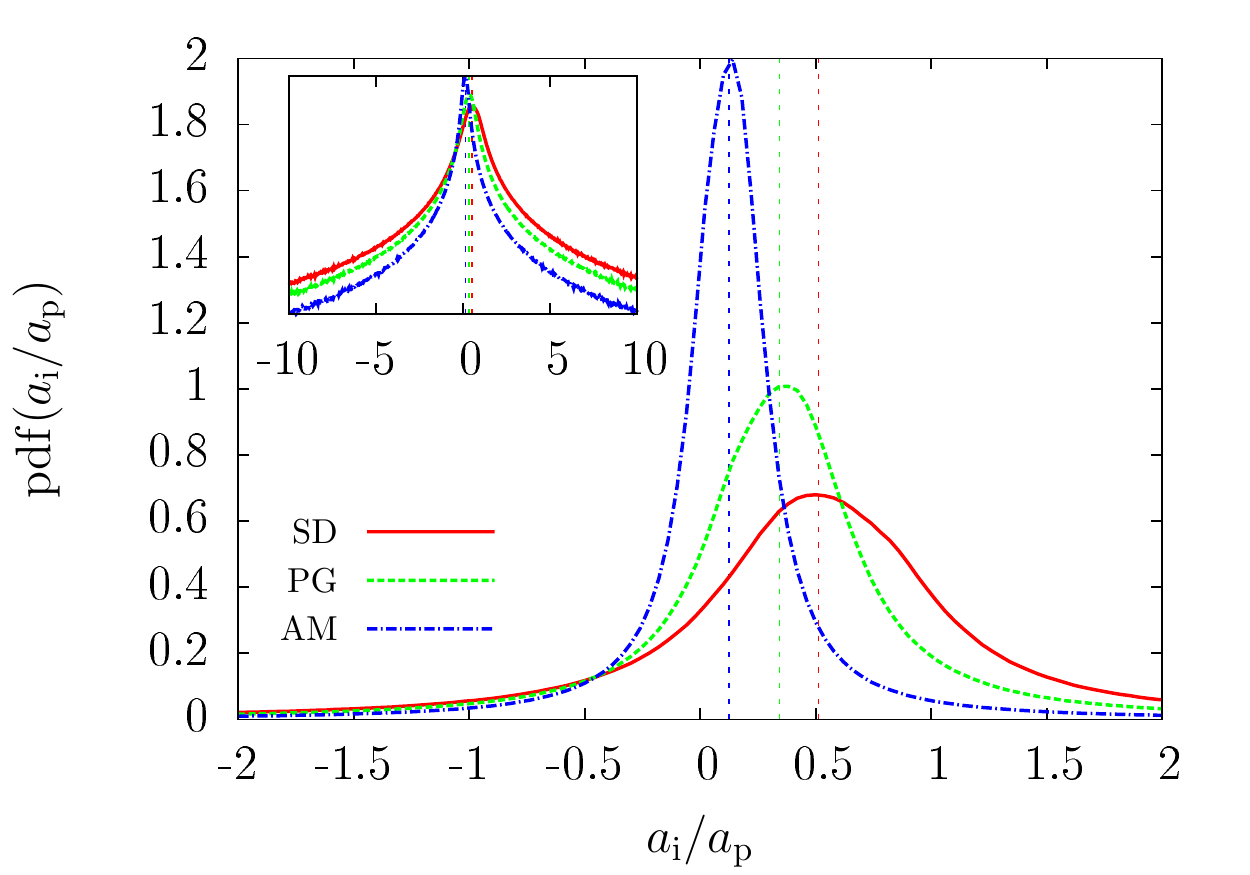}
{\scriptsize \put(-315,135){$R=10$, $St_{\mathrm{K}}=0.1$}}
{\scriptsize \put(-120,135){$R=10$, $St_{\mathrm{K}}=1.0$}}
{\put(-395,110){(a)}}
{\put(-185,110){(b)}}
\\
\includegraphics[width=.48\textwidth]{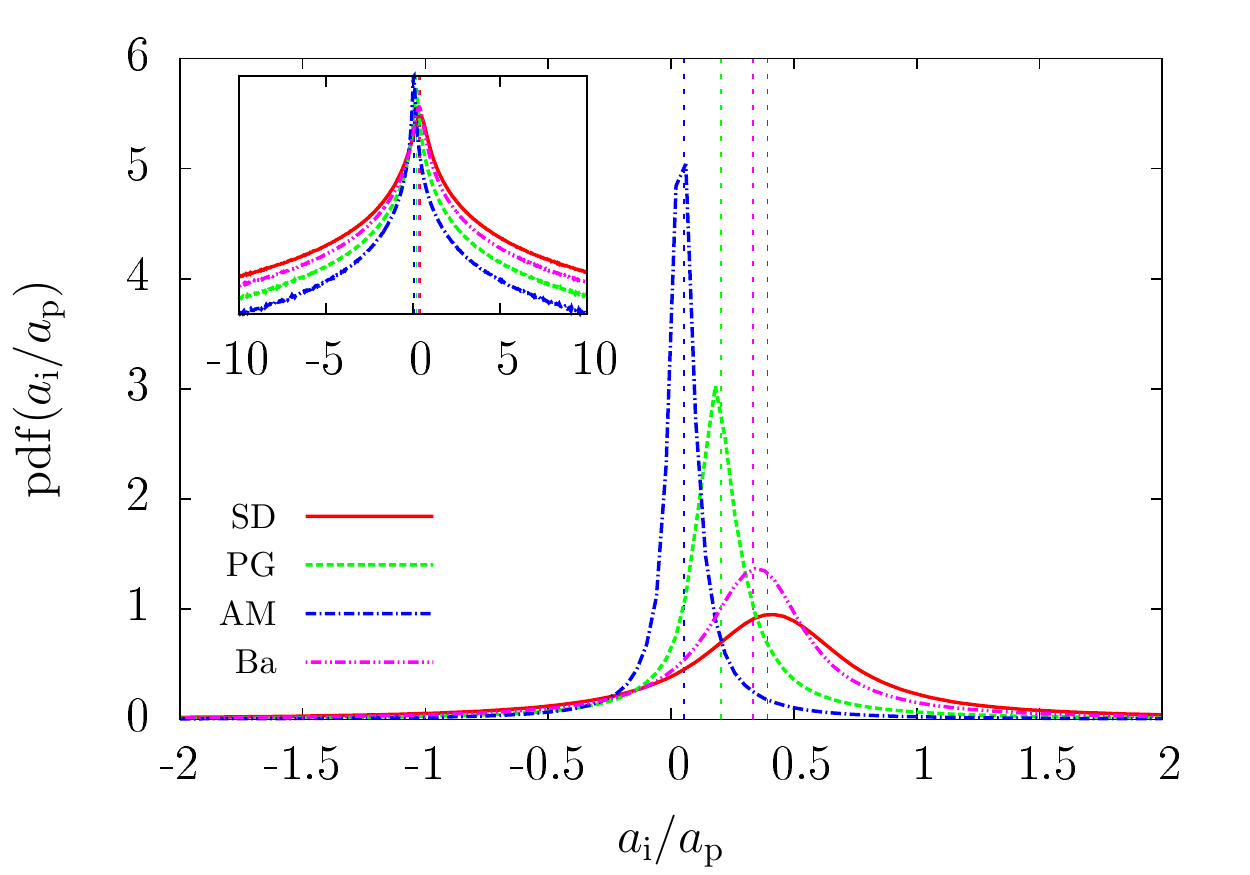}~~~~
\includegraphics[width=.48\textwidth]{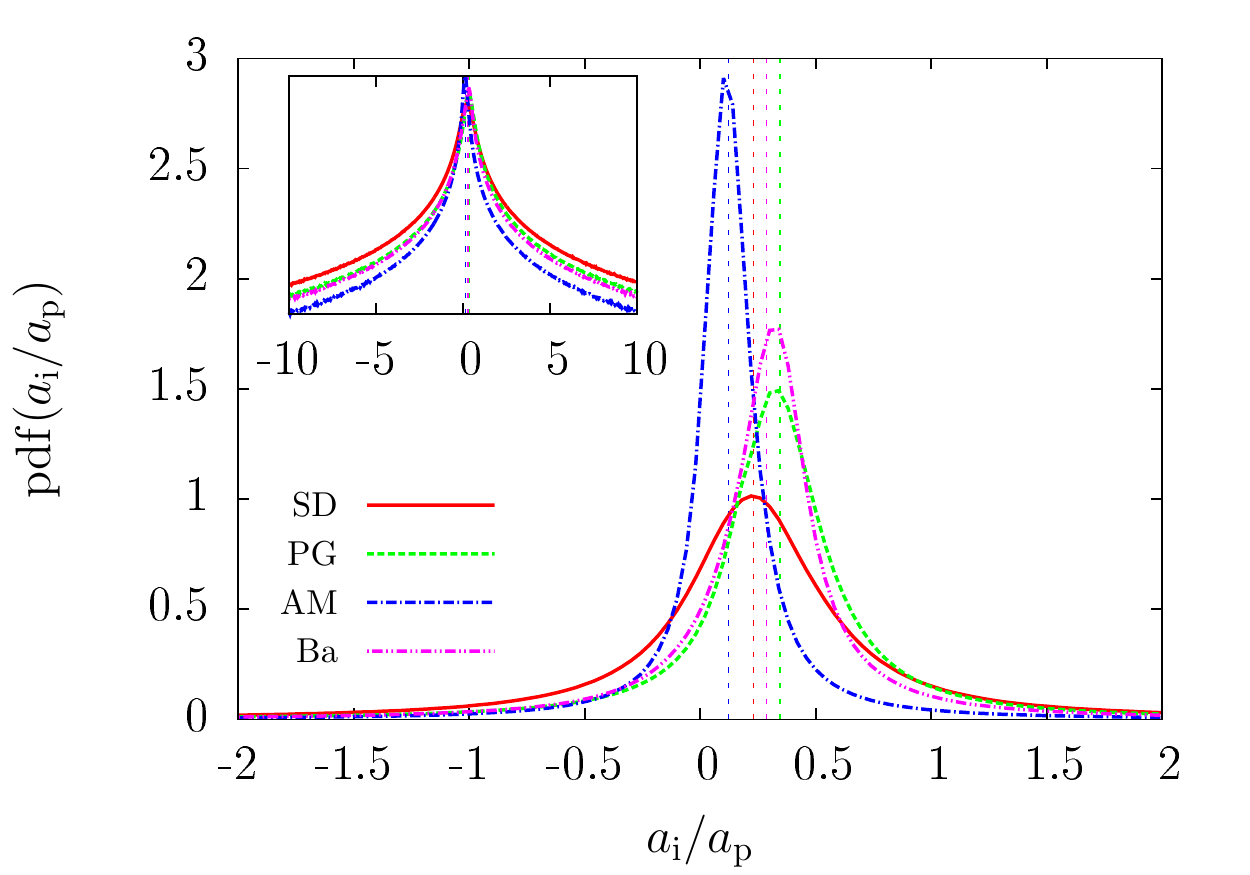}
{\put(-395,110){(c)}}
{\put(-185,110){(d)}}
\caption{P.d.f.'s of the acceleration $a_i/a_p$ for particles with $R=10$.
Data from simulations without the Basset history term in (a) and (b); and with the Basset term in (c) and (d).  {The insets are plot in lin-log axis to highlight the tails of the acceleration terms. ($\mathrm{SD}=$ Stokes Drag, $\mathrm{PG}=$ Pressure Gradient, $\mathrm{AM}=$ Added Mass, $\mathrm{Ba}=$ Basset term).}}
\label{fig:par_intermediate}
\end{figure}

\begin{figure}
\includegraphics[width=.475\textwidth]{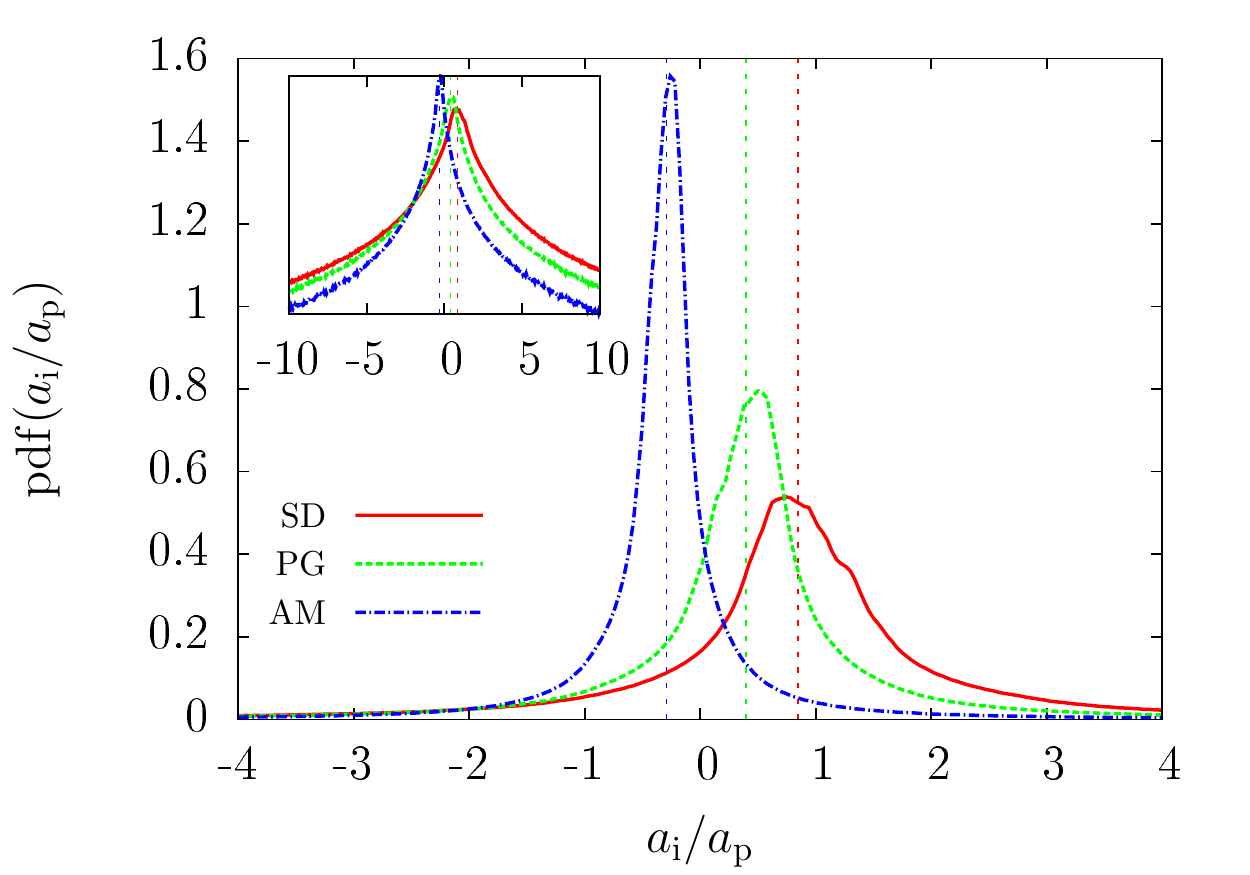}~~~~~
\includegraphics[width=.475\textwidth]{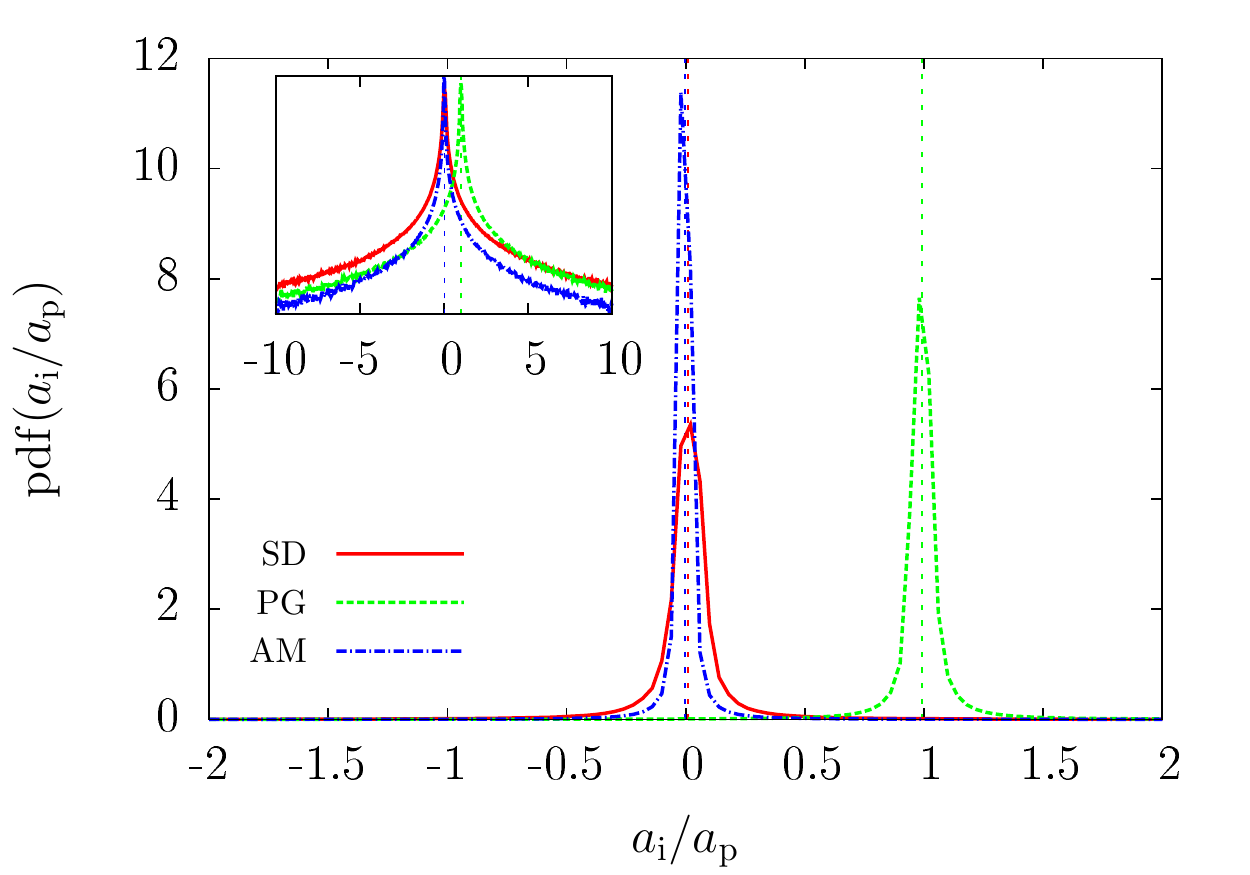}
{\scriptsize \put(-310,135){$R=1$, $St_{\mathrm{K}}=0.01$}}
{\scriptsize \put(-120,135){$R=1$, $St_{\mathrm{K}}=0.1$}}
{\put(-395,110){(a)}}
{\put(-185,110){(b)}}
\\
\includegraphics[width=.475\textwidth]{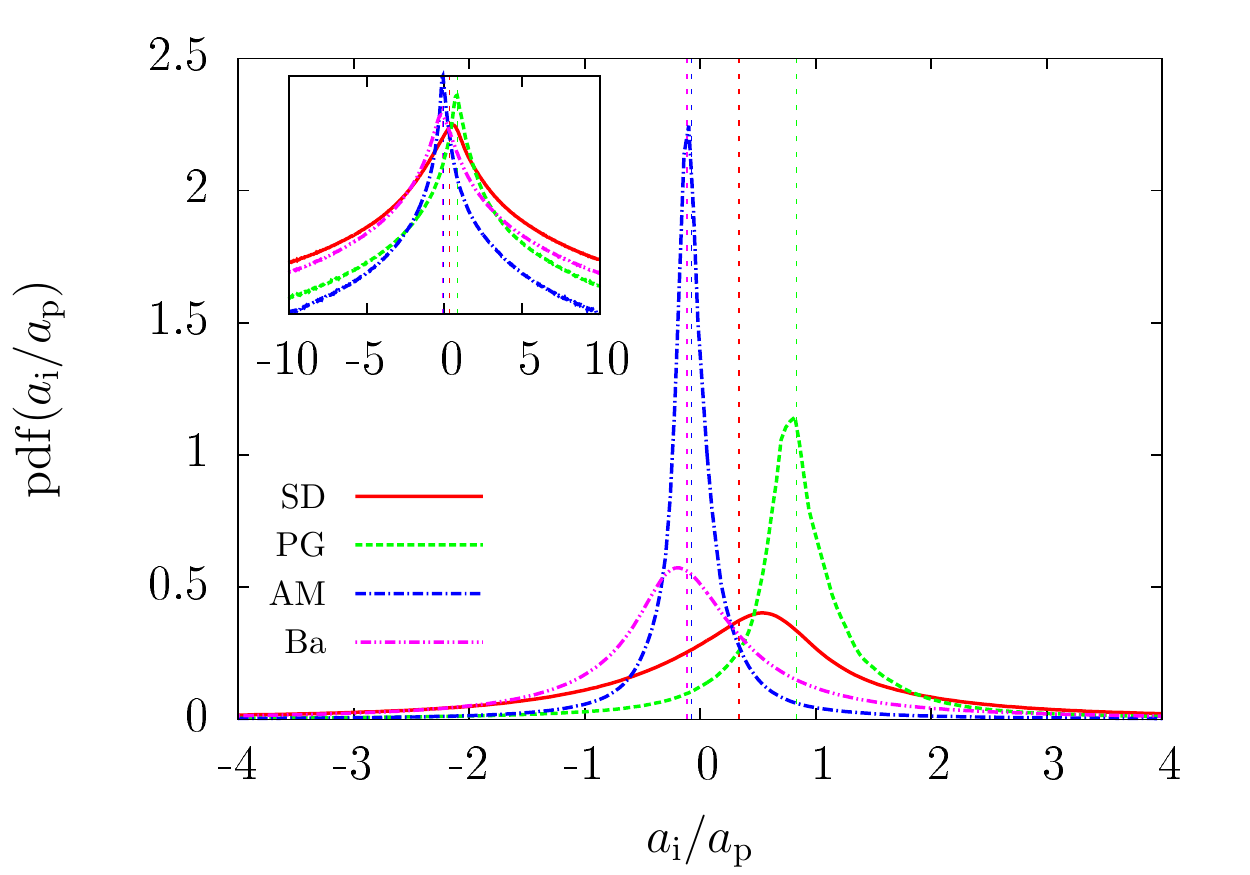}~~~~~
\includegraphics[width=.475\textwidth]{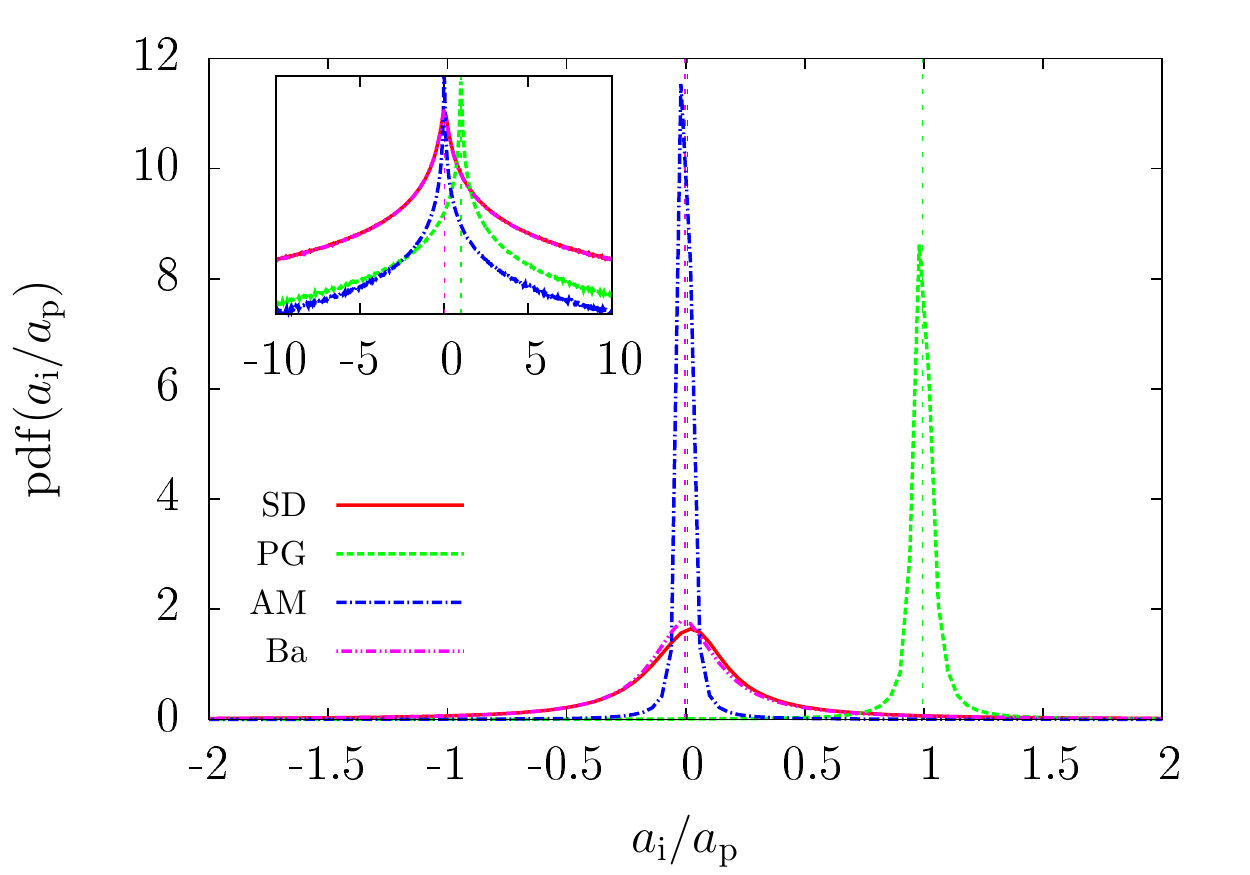}
{\put(-395,110){(c)}}
{\put(-185,110){(d)}}
\caption{P.d.f.'s of the acceleration $a_i/a_p$ for particles with $R=1$.
Data from simulations without the Basset history term in (a) and (b); and with the Basset term in (c) and (d).  {The insets are plot in lin-log axis to highlight the tails of the acceleration terms. ($\mathrm{SD}=$ Stokes Drag, $\mathrm{PG}=$ Pressure Gradient, $\mathrm{AM}=$ Added Mass, $\mathrm{Ba}=$ Basset term).}}
\label{fig:par_neutral}
\end{figure}

The behavior of particles with an intermediate density ratio, $R = 10$,  {whose clustering is most affected by Ba,} is presented in Fig.~\ref{fig:par_intermediate}.
Here, we do not find one dominant term, but the particle dynamics emerge from the contribution of the different forces, with significantly long tails  {(see the lin-log plots in the insets).}
Examining  the simulations where  {Ba} is not considered, figs.\ (a) and (b), we note that the PG becomes more and more important with respect to the SD
when increasing the Stokes number.
Most importantly, the impact of  {Ba} is relevant for all Stokes numbers considered
(see fig.\ c and d). Indeed, the average impact of the other terms, in particular SD, is strongly altered by the presence of  {Ba}. The large
difference in the clustering (RDF) shown by particles with $R=10$ and $St_K=1$ has to be ascribed to this change.  
Hence, at density ratios of the order $R\approx 10$,  {Ba} cannot be neglected to capture the
correct particle dynamics. 

Next, we report some of the results for particles of density ratio $R=1$. Although these particles do not show any clustering,
it is interesting
to document the strong alteration of the p.d.f.s of $a_i/a_p$ when  {Ba} is taken into account especially at small Stokes numbers. 
We see in Fig.~\ref{fig:par_neutral} that 
the leading term in the balance is the Stokes drag when  {Ba} is not considered, whereas it becomes 
the PG with the full model. Also for this case, we note the very long tails in the distribution of the SD.

\begin{figure}
\centering
\includegraphics[width=.48\textwidth]{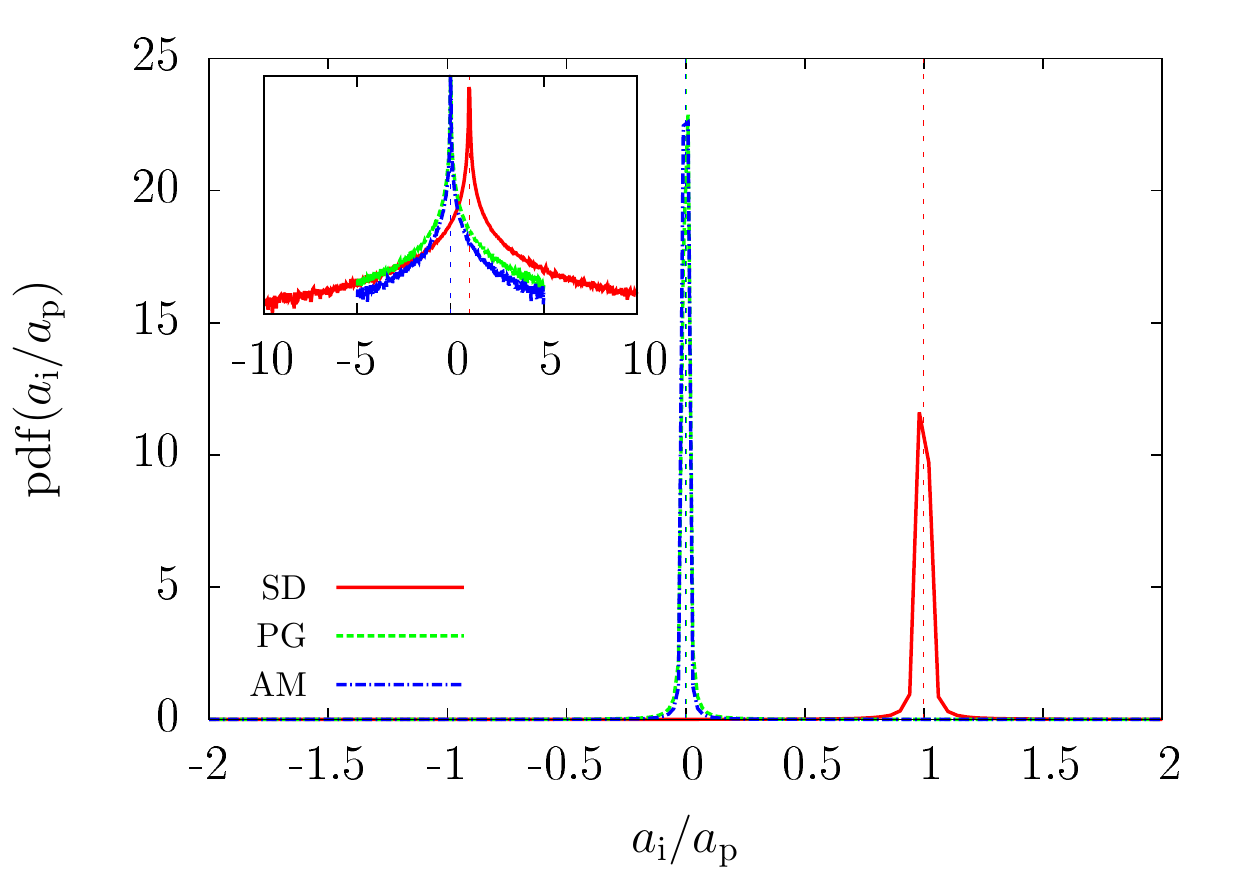}
{\scriptsize \put(-30,135){$R=1000$, $St_{\mathrm{K}}=1.0$}}
{\put(-190,110){(a)}}
{\put(5,110){(b)}}
\includegraphics[width=.48\textwidth]{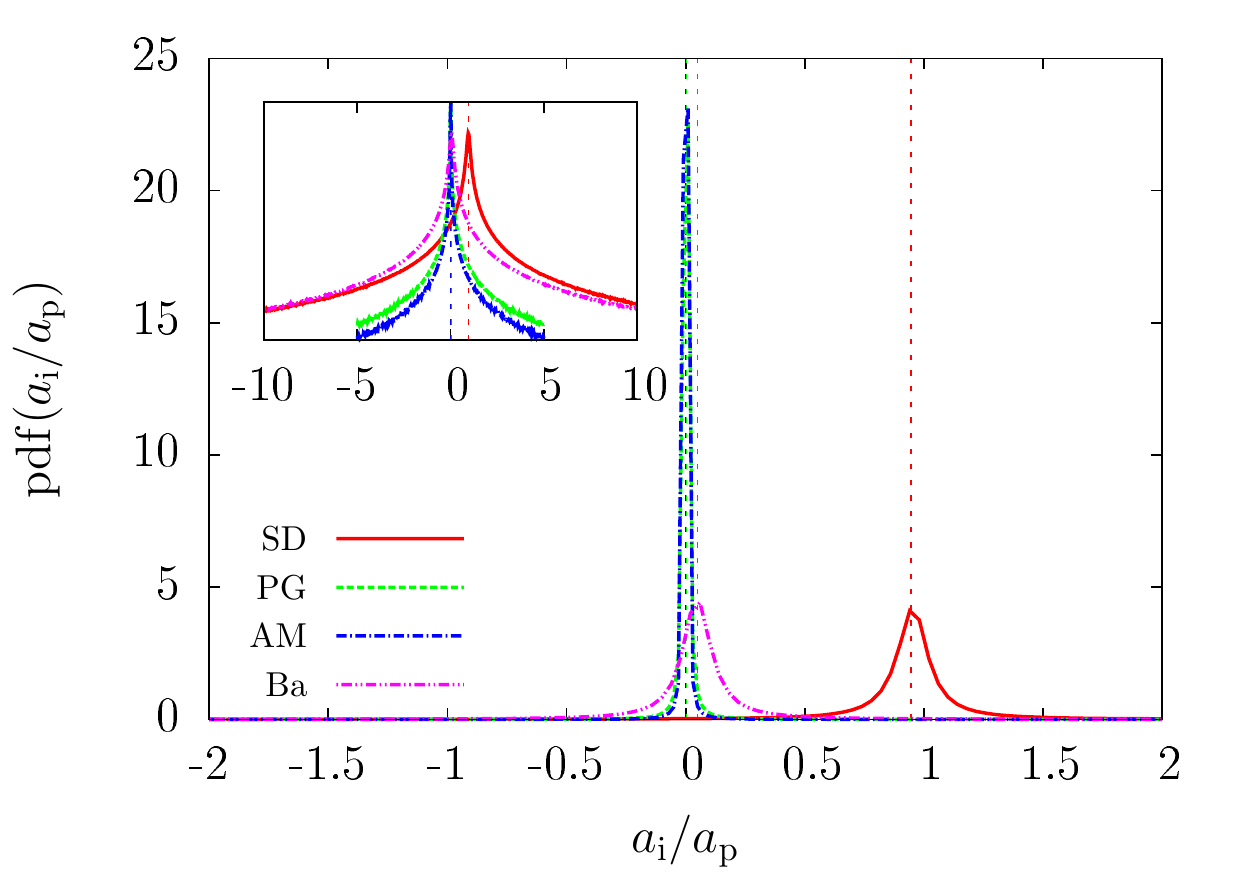}
\caption{ {P.d.f.'s of the acceleration $a_i/a_p$ for particles with $R=1000$ (heavy particles).
Data from simulations without the Basset history term in (a) and with the Basset term in (b).}
 {The insets are plot in lin-log axis to highlight the tails of the acceleration terms. ($\mathrm{SD}=$ Stokes Drag, $\mathrm{PG}=$ Pressure Gradient, $\mathrm{AM}=$ Added Mass, $\mathrm{Ba}=$ Basset term).}}
\label{fig:par_heavy}
\end{figure}

Finally, Fig.~\ref{fig:par_heavy} reports the  p.d.f.'s of the ratio $a_\mathrm{i}/a_\mathrm{p}$  {for the case with the highest density ratio here investigated, $R = 1000$, and Stokes number $St_\mathrm{K}=1$. The figure~\ref{fig:par_heavy}(a) reports simulations  without  {Ba} and shows that 
the Stokes Drag (SD) presents a narrow distribution whose average lies around 1, while the distribution of the Pressure Gradient (PG) and
the Added Mass (AM)  could be both approximated by Dirac-delta functions centered at 0. 
In other words, SD is the leading term driving the particle acceleration, in agreement with the usual assumptions in literature~\cite{armenio:2001,toschi-rev:2009}. The Basset force, however, does have an impact on the inertial particle dynamics as displayed in the right panel.
Its presence widens the p.d.f. of SD, and more importantly, moves its average to a value of about $0.9$ (vertical lines in the figure). Even at this high density ratio, Ba may influence the overall particle acceleration in an appreciable way (see discussion above about the model assumptions). 
Again, both SD and Ba exhibit long tails; rare intense events are most
influenced by the Basset history term. Particles with lower Stokes numbers show a similar behavior (not reported here).}

{\bf Final remarks.}
The present work addresses the effect of the Basset history term ( {Ba}) on the dynamics of small particles dispersed in a turbulent homogenous isotropic flow.
 The  {Ba} is found to be relevant in the
dynamics of particles with moderate density ratios, $R=1$ and $R=10$  {(i.e. sand or metal powder in water)}, 
where its presence alters the balance of the different terms that determine the particle acceleration. This has a relevant impact
 on the small-scale clustering observed at $R=10$, which is significantly reduced.

More unexpected is the impact of  {Ba} on the dynamics of particles with high density ratio, $R=1000$  {(i.e. water droplets in clouds)}: also here the clustering intensity decreases for Stokes number in the range $St_K=0.1- {1}$,  {where the diffusive model adopted starts to be questionable}. 
Examining the p.d.f.'s of the terms determining the total particle
acceleration,  {Ba} amounts to $\sim10\%$ of the total, the rest being determined by the Stokes Drag.
It is also worth noting that for all cases the p.d.f.'s of the  {Ba} show long tails, 
meaning that this force is crucial for a correct representation of rare intense events of the particle dynamics.
The effect of  {Ba} identified here could help to clarify some of the discrepancies between numerical and experimental results on particle dynamics 
that are still not fully understood, see e.g.\ Ref.~\onlinecite{calzavarini2009acceleration}. 
We believe that these findings will stimulate further investigations
on dispersed particle transport where the Basset history term will need to be considered, and new modeling efforts based on comparison with detailed particle simulations.

Computer time provided by SNIC, Swedish National Infrastructure for Computing, is gratefully acknowledged.
S.O. thanks the financial support from the PRIN 2012 project n.\ D38C13000610001 funded by the Italian Ministry of Education and the C.M. Lerici foundation. D.I. was partially funded by the Flagship Project RITMARE Ð The Italian Research for the Sea Ð funded by the Italian Ministry of Education, University, and Research within the National Research Program 2011-2013.

%


\end{document}